\documentclass[12pt,preprint]{aastex}

\begin{document}

\title{Mid-Infrared Imaging of the Protostellar Binary L1448N--IRS3(A,B)}

\author{David R. Ciardi, Jonathan P. Williams\altaffilmark{1}, Charles
M. Telesco}
\affil{211 Space Sciences Building, Department of Astronomy, University of 
Florida, Gainesville, FL 32611}
\altaffiltext{1}{Current Address: Institute for
Astronomy, 2680 Woodlawn Drive Honolulu, HI 96822}
\email{ciardi@astro.ufl.edu, jpw@ifa.hawaii.edu, telesco@astro.ufl.edu}

\author{R. Scott Fisher}
\affil{Gemini Observatory, 670 North A'ohoku Place, Hilo, HI 96720-2700}
\email{sfisher@gemini.edu}

\author{Chris Packham, Robert Pi\~na, James Radomski}
\affil{211 Space Sciences Building, Department of Astronomy, University of 
Florida, Gainesville, FL 32611}
\email{packham@astro.ufl.edu, rpina@astro.ufl.edu, jtr@astro.ufl.edu}

\slugcomment{Accepted for publication in  The Astrophysical Journal}

\begin{abstract}

Mid-infrared ($10-25$ \micron) imaging of the protostellar binary system
L1448N-IRS3(A,B) is presented. Only one source, IRS3(A), was detected at
mid-infrared wavelengths -- all of the mid-infrared emission from
IRS3(A,B) emanates from IRS3(A). The mid-infrared luminosity of IRS3(A)
is $L_{midir} = 1.3\left(\frac{d}{300\rm{pc}}\right)^2\ L_\sun$, which yields
a central source mass, depending on the mass infall rate, of $M_* = 0.2\
M_\sun \frac{10^{-6}M_\sun yr^{-1}}{\dot M}$. The envelope mass
surrounding IRS3(A) is $\sim 0.15\ M_\sun$, suggesting that the central
source and the envelope are of comparable mass. The locations of IRS3(A)
and IRS3(B) on an $M_{env} - L_{bol}$ diagram indicate that IRS3(A) and
IRS3(B) appear to be class I and class 0 protostars, respectively.

\end{abstract}

\keywords{infrared: ISM --- infrared: stars --- ISM: individual (L1448,
L1448N-IRS3) stars: formation --- stars: pre--main-sequence}

\section{Introduction}

Protostars are young stellar objects that are still in the process of
accreting the bulk of material, and with most of their luminosity being
derived from the infall of matter. Class I young stellar objects
\citep{lw84, lada87} have been viewed as protostellar candidates
\citep{als87}, but sub-mm work showed that many class I objects have
lower mass envelopes than expected, suggesting that these class I
objects may have already depleted their surrounding envelopes
\citep[e.g., ][]{bk92}. 

Class 0 sources were identified by \citet{awb93} as sources that are
more deeply embedded and associated with more powerful outflows than
``standard'' class I young stellar objects. Consequently, the class 0
sources have been proposed as the evolutionary precursors to the class I
protostars \citep{awb93, am94}. Work by \citet{bontemps96} and
\citet{saraceno96} suggested a direct evolutionary sequence from the
class 0 stage to the class I stage, with the class 0 sources
representing the youngest protostars. Observationally, the defining
characteristics of the class 0 sources are 1) a
submillimeter-to-bolometric luminosity ratio of $\gtrsim 0.5\%$, 2)
invisibility at near-infrared wavelengths, 3) a spectral energy
distribution (SED) described by a single-temperature (modified)
blackbody with a bolometric temperature of $T<70$ K, and 4) the presence
of a molecular outflow \citep{awb00, chen95, chen97}. 

However, recently \citet{jhc01} have argued that class 0 protostars are
located preferentially in higher density regions and that class I young
stellar objects are located preferentially in lower density regions and
may be of comparable evolution. Thus, the class 0 and class I sources
may not represent a true evolutionary sequence. To further complicate
the distinction between class 0 and class I objects, some class I
objects, which are viewed at high inclination, may appear ``class
0-like'' because of the high optical depth associated with viewing
the disk around class I sources (nearly) edge-on \citep{mi00}. 

To help address these issues, it would be instructive to study a set of
young stellar objects believed to be coeval and located within the same
environment, thus removing the possible ambiguity of differing initial
conditions. The binary system L1448N-IRS3 may provide such a useful
comparison.

L1448 is a globule located within the Perseus molecular cloud complex
($d\sim 300$ pc) containing an extremely young and highly collimated
bipolar outflow \citep{bachiller90}. The exciting source (L1448-mm) of
the bipolar outflow was identified by \citet{curiel90}, and has been
classified as a class 0 protostar. Located in the northwest portion of
the bipolar outflow is the infrared source L1448N-IRS3, which is
powering its own molecular outflow \citep{guilloteau92, ds95}.

L1448N-IRS3 consists of three sources which have been resolved at mm
wavelengths \citep[e.g.,][]{lmw00} and are known as L1448N-IRS3(A),
L1448N-IRS3(B), and L1448N-IRS3(C). L1448N-IRS3(C), also known as
L1448NW, is located $\sim$20\arcsec\ northwest of IRS3(A,B), and is not
considered in this paper. IRS3(A) and IRS3(B) are separated by
$\sim7\arcsec$ (2100 AU) and are considered a coeval, protostellar
binary system with a common envelope \citep{tp97, barsony98}. It has
been suggested that the strong bipolar outflow from L1448-mm may have
induced the star formation associated with L1448N-IRS3
\citep{barsony98}. 

Taken together, the IRS3(A,B) binary meets each of the observational
requirements for a class 0 protostar: 1) $L_{submm}/L_{bol} \approx 3\%$
\citep{barsony98}; 2) neither of the sources has been detected at K to a
limit of 18th magnitude \citep{tp97}; 3) the mm SED is characterized by
a temperature of 22.5 K \citep{barsony98}; and 4) there is a molecular
outflow associated with the system \citep{guilloteau92}. Detailed
modeling of IRS3(A,B) has been performed on the binary as a whole
\citep[e.g.,][]{shirley00}, and consequently, the relative evolutionary
status of the individual components of IRS3 remains undetermined.

There are differences between the two sources. At 2.7 mm, IRS3(B) is the
stronger source \citep[$F_A/F_B \approx 0.17$;][]{lmw00}, while at 2
and 6 cm, IRS3(A) dominates the emission \citep[$F_A/F_B \approx
3$;][]{curiel90}. The derived envelope masses from the resolved
(optically thin) mm data are ~0.09 M$_\sun$ and ~0.52 M$_\sun$, for
IRS3(A) and IRS3(B), respectively. While both sources are apparently
associated with outflows \citep{ds95, barsony98}, only IRS3(B) displays
signatures of mass infall, suggesting that IRS3(A) may have accumulated
most of its envelope mass \citep{tp97}. The two sources remain
unresolved for all other observations (mm wavelengths $1.3-2.6$ mm;
\citet{tca93}, and IRAS infrared wavelengths; \citet{barsony98}). The
binary system has a combined bolometric luminosity of $L\approx 10.9\
L_\sun$. \citet{barsony98} note that the IRAS infrared emission
appears in excess of that expected from the single-temperature blackbody
fit to the mm data, and they attribute the excess to shock-heated dust
associated with the outflows.

Millimeter wavelengths are most sensitive to the cool outer envelopes
surrounding protostars. Complementary to mm observations, mid-infrared
observations probe the inner regions of accretion near, or at, the
central protostar and are expected to have only a minor contribution
from the outer envelope. Thus, observations at $10-25$ \micron\ provide
a means of comparing the central protostar to the outer envelope. The
IRAS observations do not spatially resolve the components of IRS3
\citep{barsony98}. So while IRS3(A,B) is relatively bright at
mid-infrared wavelengths ($F_{12\mu{\rm m}}\approx 0.6$ Jy,
$F_{25\mu{\rm m}}\approx 6$ Jy), it is unclear how the mid-infrared
emission is distributed among the two sources and the envelope. 

In order to disentangle the mid-infrared emissions from IRS3(A),
IRS3(B), and the envelope, and to clarify the evolutionary status of the
sources, we present mid-infrared ($10-25$ \micron)
high-angular-resolution ($\sim0.5\arcsec$) observations of the
protostellar binary system. Only IRS3(A) was detected; upper limits for
the mid-infrared emission of IRS3(B) are discussed. We calculate
regional and bolometric luminosities, and estimate the central and
envelope masses. We conclude with a brief discussion of the evolutionary
status of the sources.

\section{Observations and Data Reduction}

Observations of IRS3 were made on 05 December 2000 UT using the
University of Florida mid-infrared camera OSCIR on the Gemini-North 8~m
telescope. OSCIR utilizes a $128\times 128$ Si:As blocked impurity band
detector, with a pixel scale of $0\farcs 089$ per pixel and a field of
view of $11\farcs 4 \times 11\farcs 4$ on the Gemini-North telescope.
IRS3 was observed in four filter passbands (N, IHW18, Q3, \& H25) using
a standard chop-nod sequence with a 30\arcsec\ chop throw in
declination. The weather was mostly clear with some high cirrus and
typical seeing of $<0\farcs5$. A summary of the filters, frametimes,
total integration time per filter, and associated airmasses is given in
Table 1.

To ensure accurate pointing, the telescope was first centered on the
optical guide star located $\sim$5\arcmin\ from IRS3, and the telescope
was offset to a position located directly between IRS3(A) and IRS3(B).
The initial source acquisition was made with the H25 filter ($\lambda_c
\sim 24$ \micron). The flux calibration was obtained by observing
$\beta$ And just prior to the science observations and $\alpha$ Tau just
after the science observations.

IRS3(A) was immediately apparent in the real-time OSCIR data acquisition
quick-look display. To remove the possibility of pointing errors and
ambiguity of the identity of the detected source, the offset pointing
was confirmed and a three-position mosaic was made with the H25 filter,
with 5\arcsec\ steps to the northwest. No detection of IRS3(B) or any
other source was made. The data were reduced with custom-written IDL
routines for the OSCIR data format. The images were smoothed with an
$0\farcs4$ gaussian, and standard aperture photometry was performed on
the final reduced co-added frames using an IDL version of PHOT, with an
aperture diameter of $0\farcs75$. The H25 (24 \micron) mosaic is
presented in Figure 1.

\section{Discussion}

The primary result, as indicated by Figure 1, is that {\it all} of the
mid-infrared flux observed from L1448N-IRS3 emanates from IRS3(A). All
of the detected flux arises within a region of diameter $0\farcs75$;
i.e., the mid-infrared emission comes from a region located {\em no
farther} than $\sim 100$ AU, in projection, from the central source.
\citet{barsony98} argued that the infrared emission (12 \& 25 \micron),
in excess of the envelope contribution, may be the result of
shock-heated dust resulting from interactions between the local dust and
the outflows. No mid-infrared emission is observed toward IRS3(B), which
powers its own outflow, or in the general environment, suggesting that
the mid-infrared emission of IRS3 is associated entirely with the
accretion emission of IRS3(A).

Figure 1 does show a hint of extended structure at 24 \micron\ which is
perpendicular to the direction of the H$_2$ outflow emission from IRS3
\citep{davis94}, perhaps suggestive of the accretion structure around
IRS3(A). The other mid-infrared images were searched for evidence of
this structure, and nothing conclusive was found. However, the
integration time for the 24 \micron\ image is 3 times longer than the
integrations times in other filters. At present, it is unclear if the
observed structure is real.

\subsection{Mid-Infrared Spectral Energy Distribution}

The photometry for IRS3(A) is presented in Table 2, and is in agreement
with the IRAS HIRES flux densities reported in \citet{barsony98}. While
IRS3(B) was not detected, 3$\sigma$ upper limits were calculated based
upon the statistical uncertainties of photometry for IRS3(A). In
addition to the OSCIR data presented here, the ISO database was
searched, and a 14 \micron\ ISOCAM image was extracted. ISO clearly
detected IRS3, but did not resolve the two sources. Here we assume that
the ISO 14 \micron\ emission is contributed solely by IRS3(A), as is the
case for our mid-infrared photometry. Using a 20\arcsec\ aperture, the
14 \micron\ flux density is measured to be $F_{14\mu{\rm m}} = 0.5 \pm
0.1$ Jy. The OSCIR and ISO mid-infrared SED of IRS3(A) is displayed in
Figure 2.

To understand the temperatures associated with the mid-infrared
emission, a single temperature blackbody was fitted to the dataset. A
$T=148\pm6$ K blackbody\footnote{Uncertainties for the blackbody fitting
were estimated via a Monte Carlo simulation where the data points were
randomly adjusted by their individual uncertainties, and the data were
re-fitted. The simulation was performed 5000 times, and the final
uncertainties were estimated from the standard deviations of the best
fits.} characterizes the overall $10-25$ \micron\ colors, but is
generally a poor fit to the dataset with a $\chi^2 \sim 11$. In the SED
plot (Fig. 2), the $T\sim148$ K blackbody curve is shown, and it clearly
does not represent accurately the distribution of the mid-infrared
fluxes. 

If the blackbody fitting is restricted to the $\lambda > 18$ \micron\
data, a single temperature $T=82\pm9$ K blackbody fits the longer
wavelength data very well ($\chi^2_\nu \sim 0.3$), but significantly
underestimates the 10 \& 14 \micron\ fluxes (see Fig. 2). We also
considered a single-temperature modified blackbody function of the form
$S_\nu = B_\nu(T_d)(1-e^{- \tau_\nu})\Omega$, where $B_\nu(T_d)$ is the
Planck function, $\tau_\nu$ is the optical depth with a $\nu^\beta$
dependency, and $\Omega$ is the solid angle. However, we did not find a
reasonable fit to the dataset. In order to parameterize the
observations, multiple source temperatures are necessary.

A two temperature blackbody function of the form $S_\nu(T_1, T_2) =
s\cdot [f \cdot B_\nu(T_1) + (1-f) \cdot B_\nu(T_2)]$, where $T_1$ and
$T_2$ are the blackbody temperatures, $s$ is the scaling factor, and $f$
is the fractional contribution of the primary blackbody component, was
fitted to the mid-infrared data. In general, the scaling factor $s$ is
a multiplicative combination of the source solid angle and the radiative
transfer function $(1-e^{- \tau})$. The relative contributions $f$ and
$(1-f)$ may represent source size and/or optical depth differences. As a
final caveat, the broad N-band 10 \micron\ filter and the ISO 14
\micron\ filter span the silicate feature. Without narrow-band
photometry or low-resolution spectroscopy, it is not possible to
determine how the silicate feature contaminates the photometry. Since
the silicate feature, if present, is likely to be in absorption for such
young and highly embedded sources, the 10 and 14 \micron\ fluxes may
underestimate the overall mid-infrared continuum level.

The best fit parameters for the dual-blackbody model consist of two
distinct temperatures of $T_1 = 71 \pm 8$ K and $T_2 = 240 \pm 59$ K
with relative contributions of $f=99.98$\% and $(1-f)=0.02$\%, and a
scale factor of $s=9.4\times10^{-12}$. The resultant reduced chi-square
is $\chi^2_\nu = 0.2$. The summed curve and the individual components of
the fit are shown in Fig. 2.

A two temperature blackbody model is obviously simplistic given that the
temperature distribution is likely to be continuous with, for example, a
radial powerlaw fall-off of the form $T(r) \propto r^{-0.4}$
\citep{kch93, tca93}. However, it does characterize the emission and
emphasizes the regions from where the mid-infrared emission arises. Dust
temperatures of $T \sim 50-100$ K are located at radii of $r \approx 10
- 100$ AU \citep[e.g., ][]{kch93}, consistent with the $0\farcs 75$
($<100$ AU) source size of IRS3(A) (Fig. 1). Thus, the mid-infrared
observations are probing directly the zone of accretion near, or at, the
central protostar.

The detected mid-infrared radiation provides a means of estimating the
central accretion luminosity. Summing over the dual blackbody curve from
$5-30$ \micron, we estimate the mid-infrared luminosity to be $L_{midir}
= 1.3\left(\frac{d}{300\rm{pc}}\right)^2\ L_\sun$. If all the mid-infrared
luminosity is a result of gravitational infall, then we can estimate a
central source mass from the relation $L = \frac{G\dot{M} M_*}{R_*}$,
where $\dot{M}$ is the mass infall rate, $R_*$ is the source size, and
$M_*$ is the source mass. Using a standard $R_* = 3\ R_\sun$
protostellar radius \citep{sst80} and typical mass accretion rates of
$\dot{M} = 10^{-5} - 10^{-6}\ M_\sun\ {\rm yr}^{-1}$ \citep{kch93}, we
estimate the central mass to be $M_* = 0.02 - 0.2\ M_\sun$. 

\subsection{Overall Spectral Energy Distribution}

At $\lambda = 2.7$ mm, IRS3(A) and IRS3(B) have been spatially resolved
as two distinct sources, but with a common envelope \citep{lmw00}. The
2.7 mm flux densities, integrated over $\sim 2.5\arcsec$ boxes, for
IRS3(A) and IRS3(B) are $F_{2.7mm}=0.019\pm0.004$ Jy and
$F_{2.7mm}=0.116\pm0.01$ Jy, respectively. At far-infrared wavelengths,
the binary remains unresolved, with IRAS flux densities of $F_{60\mu m}
= 29 \pm 6$ Jy and $F_{100\mu m} = 89 \pm 19$ Jy \citep{barsony98}. In
Figure 3, we present the overall spectral energy distribution of
IRS3(A). 

The modified blackbody fit to the IRS3(A,B) common envelope
\citep{barsony98}, scaled to the 2.7 mm flux of IRS3(A), is shown in
Figure 3. The cold ($T=22.5$ K) blackbody fit to the envelope clearly
does not predict the observed mid-infrared luminosity. Unlike the
prototypical class 0 sources (e.g., VLA 1623), the cold envelope
emission does not dominate the overall SED of IRS3(A)
\citep[e.g.,][]{awb00}.

Using spherical infalling envelope models \citep{tsc84, kch93,
hartmann98}, the envelope mass and bolometric luminosity of IRS3(A) can
be estimated. The models (provided courtesy N. Calvet) have a density at
$r=1$ AU of $\rho_1 = 7.5 \times 10^{-14}$ g cm$^{-3}$, corresponding to
an envelope infall rate of $\dot{M} \approx 10^{-5} M_\sun\ {\rm
yr}^{-1}$. Models with centrifugal radii of $r_c = 50$ AU and 150 AU
were tested. The centrifugal radius $r_c$ represents the radius at which
the spherical envelope departs from free-fall ($\rho \propto r^{-3/2}$)
as material falls onto the disk ($\rho \propto r^{-1/2}$). The models
have an inclination of $i \sim 60^{\circ}$ and a luminosity of 1
$L_\sun$. A factor of 1.6 was needed to scale the models to the 2.7 mm
flux, indicating that IRS3(A) is more luminous than 1 $L_\sun$ (in
agreement with the measured $L_{midir} = 1.3\ L_\sun$). The model SEDs
are shown in Figure 3.

The shape of the model SED and the depth of the silicate features are
dependent upon the source luminosity, density, centrifugal radius, and
source inclination. Detailed modeling of the SED cannot provide a unique
solution without a much more complete SED for IRS3(A). However,
representative models can qualitatively parameterize the luminosity and
the density of the protostar. 

A model with $\rho_1 \approx 10^{-13}$ g cm$^{-3}$ and $\dot{M} \approx
10^{-5}$ M$_\sun\ {\rm yr}^{-1}$ provides a reasonable fit to the
overall SED of IRS3(A) (see Figure 3). For this particular model density
($\rho_1$), the model with the smaller centrifugal radius (50 AU) best
describes the data. If the density $\rho_1$ is decreased by an order of
magnitude, then the models predict an infrared luminosity $10-100$ times
larger than that observed. If the density is increased by an order of
magnitude, the models predict an infrared luminosity $10-100$ times
lower than that observed \citep[e.g.,][]{kch93, hartmann98}. These model
densities and infall rates are more like those associated with the
models for class I protostars such as L1551-IRS5 \citep[e.g.,][]{kch93,
hartmann98}. \citet{jhc01} found that the true class 0 sources (VLA 1623
\& HH24 MMS) require models with infall rates an order of magnitude
higher than required for IRS3(A).

The bolometric luminosity of IRS3(A), estimated from the luminosity of
the scaled infall model, is $L_{bol} \approx 1.6\ L_\sun$, and the submm
luminosity ($\lambda > 350$ \micron) is $L_{submm} \approx 10^{-4}
L_\sun$. Thus, $L_{submm}/L_{bol} \sim 0.01$\% is below the
observational minimum for a class 0 protostar. The model envelope mass
($r > r_c$) is $M_{env} \approx 0.15\ M_\sun$. Finally, the modeling
suggests that the majority, if not all, of the far-infrared emission of
IRS3(A,B) is also produced by IRS3(A). 

\subsection{Classification of IRS3} 

A key goal of the SED analyses is to clarify the evolutionary status of
the IRS3 components. Class 0 to class I evolution is characterized by
the transition from $M_{env} > M_*$ to $M_{env} < M_*$. The envelope
mass for IRS3(A) was found to be $M_{env}\approx 0.15\ M_\sun$, and
depending upon the assumed infall rate $\dot M$, the IRS3(A) central
source mass is estimated to be $M_* \sim 0.02 - 0.2\ M_\sun$. Thus, the
envelope and central star masses are comparable to each other.

\citet{bontemps96} established an empirical diagnostic $M_{env} -
L_{bol}$ diagram that appears to naturally separate confirmed class 0
sources from confirmed class I sources \citep[see also ][]{awb00}. We
have reproduced that diagram here (Figure 4) and have indicated the
position of IRS3(A). The minimum $L_{bol}$ is from the mid-infrared
luminosity estimate (1.3 $L_\sun$) and the maximum is from the scaled
infall model (1.6 $L_\sun$). 

IRS3(A) falls squarely within the midst of the confirmed class I
objects, and is fully separated from the class 0 protostars. The
position of IRS3(A) in the $M_{env} - L_{bol}$ diagram, combined with
the relative mass estimates from above, suggests that IRS3(A) may be
more evolved than a class 0 protostar. This may also explain why infall
appears associated with IRS3(B) and not IRS3(A) \citep{tp97}. The fact
that IRS3(A) is invisible at near-infrared wavelengths may have more to
do with the common envelope than the actual evolutionary status of
IRS3(A). Perhaps IRS3(A) marks a transition between the class 0
protostar and the class I young stellar object.

We can also estimate the classification of IRS3(B) from the $M_{env} -
L_{bol}$ diagram. The total luminosity of the binary is $L_{bol} = 10.9\
L_\sun$ \citep{barsony98}. A range of $L_{bol} \approx 9.3 - 9.6 \
L_\sun$ can be assigned to IRS3(B), if we associate $1.3 - 1.6\ L_\sun$
to IRS3(A). The envelope mass, as determined from the 2.7 mm flux
density, is $M_{env}\approx 0.5 M_\sun$. On the $M_{env} - L_{bol}$
diagram, these values place IRS3(B) among the class 0 sources. If
IRS3(A) and IRS3(B) are indeed coeval, then the apparent classification
difference between the two sources may yield information about how
binary stars evolve and how they interact with each other and/or their
environments.

\section{Summary}

We have obtained high angular resolution ($\sim0\farcs5$) $10-25$
\micron\ imaging of the protostellar binary system L1448N-IRS3(A,B).
These observations mark the first time the sources have been spatially
resolved at wavelengths $\lambda < 2.7$ mm. The binary, as a whole, had
been previously classified as a class 0 protostar. The primary result of
the work is that only one source, IRS3(A), was detected at mid-infrared
wavelengths. That is, all of the mid-infrared emission from IRS3(A,B)
emanates from IRS3(A). The mid-infrared luminosity of IRS3(A) is
$L_{midir} = 1.3\ L_\sun$. The central source mass ($0.02 - 0.2\
M_\sun$) of IRS3(A) is comparable to its envelope mass ($0.15\ M_\sun$).

IRS3(A), with a $L_{submm}/L_{bol} \sim 0.01$\% and an infall rate of
$10^{-5}$ M$_\sun\ {\rm yr}^{-1}$, may be more akin to the class I
protostars. The common envelope surrounding IRS3(A) and IRS3(B) may skew
the appearance of IRS3(A). Perhaps IRS3(A) is viewed at a particular
angle that gives it the appearance of both a class 0 and class I source,
or perhaps IRS3(A) is a transition object between the two classes.
IRS3(B), invisible at near and mid-infrared wavelengths, appears to
possess a SED dominated by the cold envelope, indicating that IRS3(B)
may be a true class 0 protostar. Spatially resolved far-infrared
observations ($\lambda = 30 - 300$ \micron) with SIRTF, coupled with
detailed modeling, are needed to determine the relative contributions of
IRS3(A) and IRS3(B) to the emission near the peak of the spectral energy
distributions, and the relative evolutionary status of the two sources.

\acknowledgments

The authors would like thank the Gemini staff for their help and
outstanding support in making OSCIR a success on Gemini-North. The UF
team would especially like to thank Chris Carter, who endured some long
nights at the telescope while the astronomers worked - although the wee
dram or two at the end was probably worth it. D.R.C. would like to thank
N. Calvet for her kind and helpful discussions and for providing the
protostar models in numerical format. The authors would like to thank
the referee, whose comments led to a better paper. J.P.W. acknowledges
support by NSF grant AST-0134739. C.M.T. acknowledges support by NSF
grant AST-0098392.

Based on observations obtained at the Gemini Observatory, which is
operated by the Association of Universities for Research in Astronomy,
Inc., under a cooperative agreement with the NSF on behalf of the Gemini
partnership: the National Science Foundation (United States), the
Particle Physics and Astronomy Research Council (United Kingdom), the
National Research Council (Canada), CONICYT (Chile), the Australian
Research Council (Australia), CNPq (Brazil) and CONICET (Argentina).  This
paper is based on observations obtained with the mid-infrared camera
OSCIR, developed by the University of Florida with support from the
National Aeronautics and Space Administration, and operated jointly by
Gemini and the University of Florida Infrared Astrophysics Group.

\clearpage

\begin{figure}[b]
\plotone{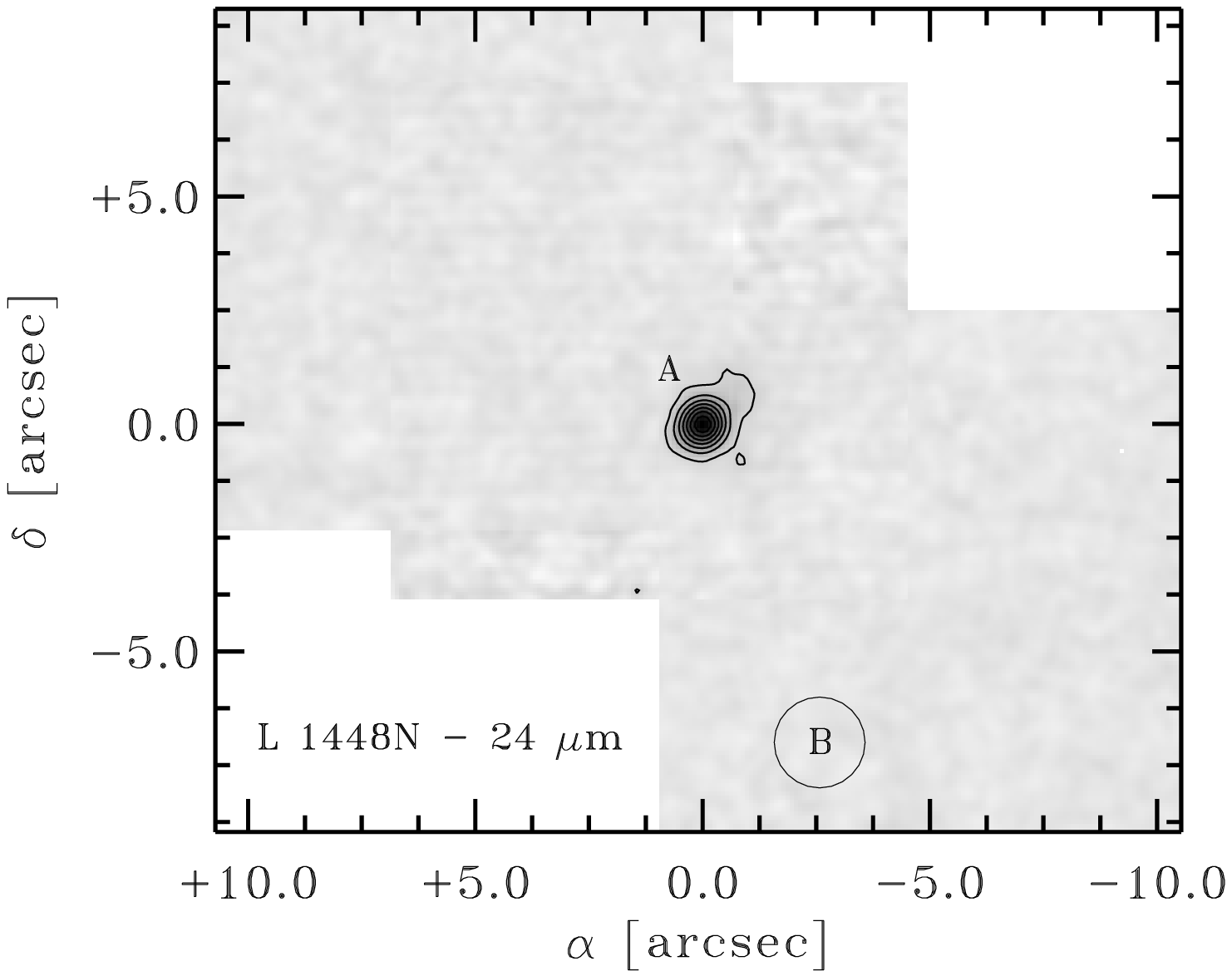} 

\figcaption{24 \micron\ image of L~1448N--IRS3, smoothed with an
$0\farcs4$ gaussian. Only IRS3(A) is detected. The position of IRS3(B)
is marked by the open circle. The contours start at 0.6 Jy/arcsec$^2$
and are stepped by 1.0 Jy/arcsec$^2$. $(0,0) =\alpha (J2000) = 03h\ 25m\
36.5s,\ \delta (J2000)= 30^\circ\ 45\arcmin\ 21\arcsec$.}

\end{figure}

\begin{figure}[b]
\epsscale{0.75} 
\plotone{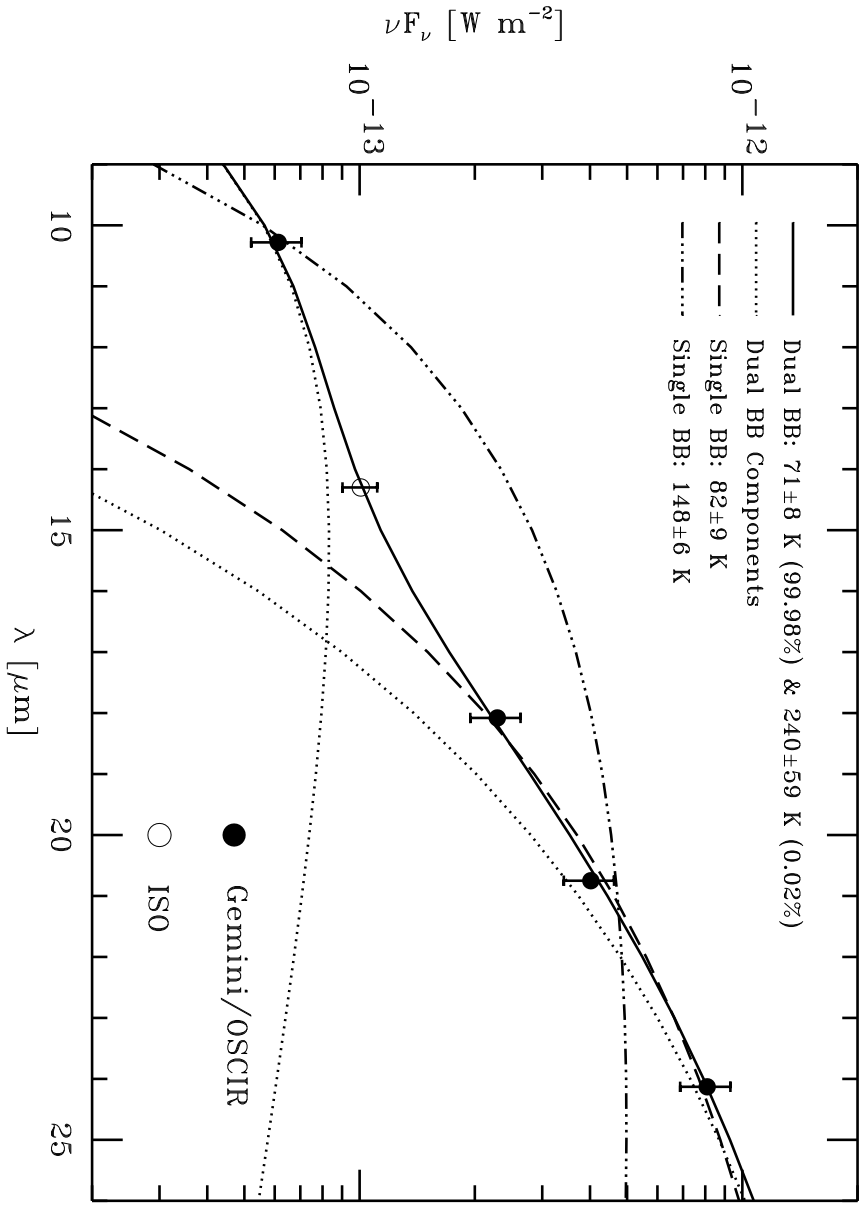}
\figcaption{Mid-infrared spectral energy distribution for IRS3(A). The
148 K single blackbody curve was fitted to all of the mid-infrared data,
and the 82 K single blackbody curve was fitted to only the OSCIR IHW18,
Q3, \& H25 filter points. The dual blackbody curve was fitted to all of
the mid-infrared data; the individual components of the dual blackbody
are also shown.}
\end{figure}

\begin{figure}[bh]
\epsscale{0.65} 
\plotone{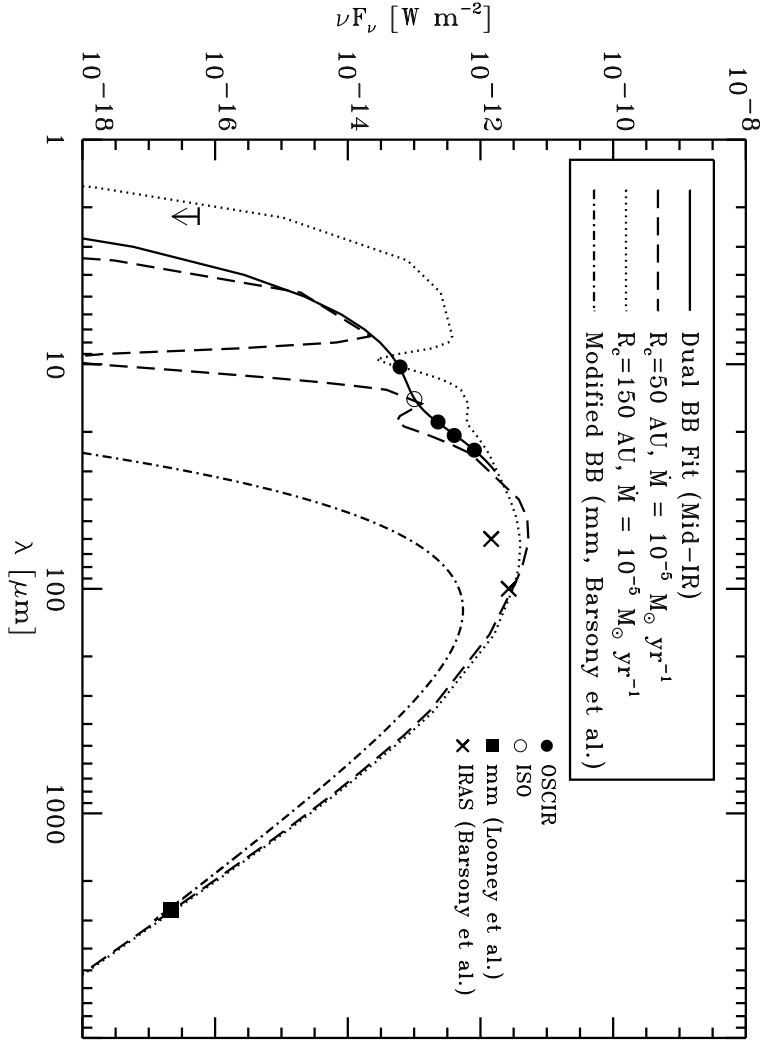}

\figcaption{Spectral energy distribution for IRS3(A), including the $K>
18$ mag upper limit \citep{tp97}. The 60 \& 100 \micron\ IRAS data
represent the total (unresolved) far-infrared flux of the binary. The
dual blackbody curve is from Figure 2, and the modified blackbody
function, scaled to the $\lambda = 2.7$ mm flux, is from
\citet{barsony98}. The dashed and dotted curves represent spherical
infalling envelope models (courtesy N. Calvet).}

\end{figure}

\begin{figure}[b]
\epsscale{1.0} 
\plotone{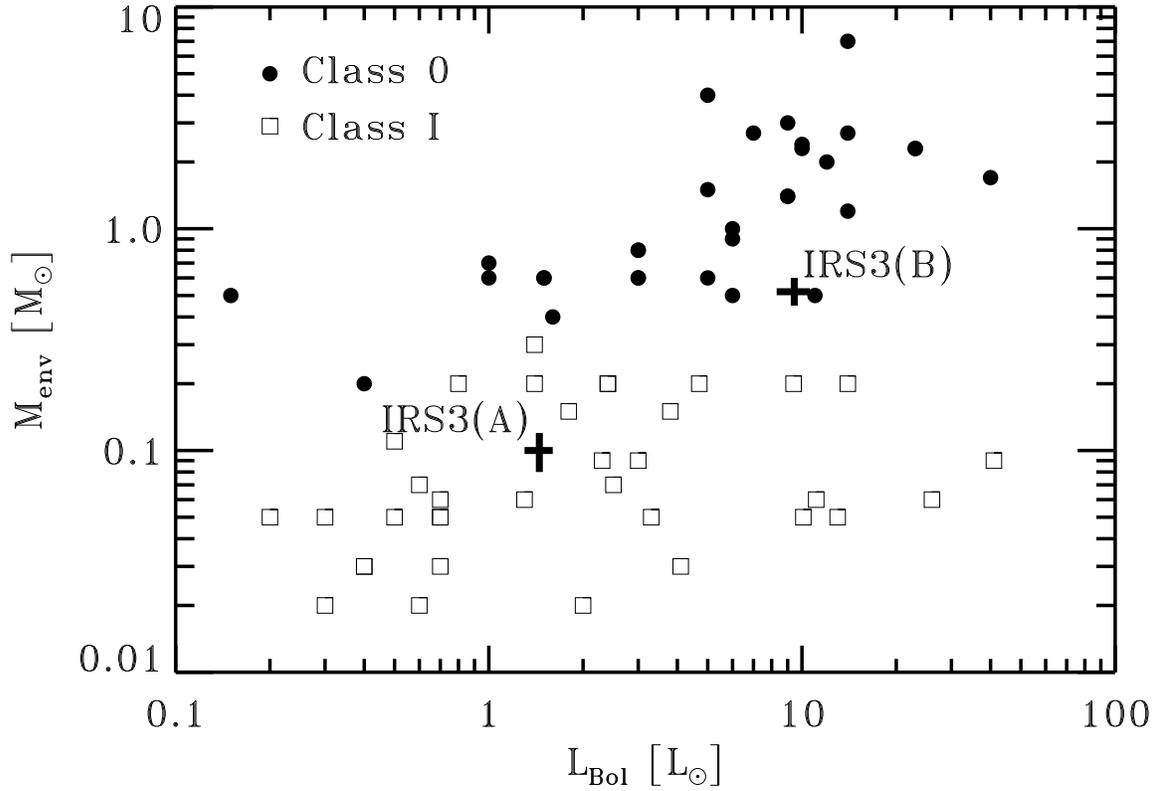}
\figcaption{Envelope mass versus bolometric luminosity is plotted for a
set of Class 0 and Class I protostars. Figure and data adapted from
\citet{bontemps96} and \citet{awb00}. The positions of IRS3(A) \& IRS3(B)
are also shown.}
\end{figure}

\clearpage
\begin{deluxetable}{cccccc}
\tablecolumns{6} \tablewidth{0pc} \tablecaption{Summary of Observations}
\tablehead{ \colhead{Filter} & \colhead{$\lambda_c$} & \colhead{$\Delta\lambda$} 
& \colhead{Frametime} & \colhead{On-Source} & \colhead{Airmass}\\
\colhead{} & \colhead{(\micron)} & \colhead{(\micron)} & \colhead{(ms)} 
& \colhead{(min)} & \colhead{} } 
\startdata
N & 10.38 & 5.23 & 7 & 2 & 1.5\\
IHW18 & 18.08 & 1.65 & 13 & 2 & 1.4\\
Q3 & 20.75 & 1.65 & 13 & 2 & 1.3\\
H25 & 24.13 & 3.6 & 7 & 6\tablenotemark{a} & 1.2 -- 1.3\\
\enddata
\tablenotetext{a}{Summation of 3-position mosaic with 2 minutes per
position.}
\end{deluxetable}

\begin{deluxetable}{ccc}
\tablecolumns{3} \tablewidth{0pc} \tablecaption{IRS3 Mid-Infrared Photometry}
\tablehead{ \colhead{Filter} & \colhead{IRS3(A)} & 
\colhead{IRS3(B)\tablenotemark{a}}\\ 
\colhead{} & \colhead{$F_\nu$ (Jy)} & \colhead{$F_\nu$ (Jy)}
 } 
\startdata
N & $0.21\pm0.01$ & $<0.03$\\
IHW18 & $1.38\pm0.02$ & $<0.06$\\
Q3 & $2.78\pm0.04$ & $<0.12$\\
H25 & $6.51\pm0.08$ & $<0.24$\\
\enddata
\tablenotetext{a}{$3\sigma$ upper limits.}
\end{deluxetable}

\end{document}